\documentclass[twocolumn,prstper,superscriptaddress,aps,10pt]{revtex4-1} 

\usepackage{amsmath}  
\usepackage{amsfonts} 
\usepackage{graphicx} 
\usepackage{hyperref}
\usepackage{enumitem}
\usepackage{minibox}
\usepackage{color}
\usepackage{mdframed}
\usepackage{booktabs}
\usepackage{multirow}
\usepackage{array}

\begin{document}

\title{Assessing Learning Outcomes in Middle-Division Classical Mechanics: The Colorado Classical Mechanics/Math Methods Instrument}

\author{Marcos D. Caballero}
\email{Corresponding Author: caballero@pa.msu.edu}
\affiliation{Department of Physics and Astronomy, Michigan State University, East Lansing, MI 48824, USA}
\author{Leanne Doughty}
\affiliation{School of Education and Human Development, University of Colorado Denver, Denver, CO 80204, USA}
\author{Anna M. Turnbull}
\affiliation{Lyman Briggs Colloege \& Department of Physics and Astronomy, Michigan State University, East Lansing, MI 48824, USA}
\author{Rachel E. Pepper}
\affiliation{Department of Physics, University of Puget Sound, Tacoma, WA 98416, USA}
\author{Steven J. Pollock}
\affiliation{Department of Physics, University of Colorado Boulder, Boulder, CO 80309, USA}

\date{\today}

\begin{abstract}
Reliable and validated assessments of introductory physics have been instrumental in driving curricular and pedagogical reforms that lead to improved student learning. 
As part of an effort to systematically improve our sophomore-level Classical Mechanics and Math Methods course (CM 1) at CU Boulder, we have developed a tool to assess student learning of CM 1 concepts in the upper-division. 
The Colorado Classical Mechanics/Math Methods Instrument (CCMI) builds on faculty consensus learning goals and systematic observations of student difficulties. 
The result is a 9-question open-ended post-test that probes student learning in the first half of a two-semester classical mechanics / math methods sequence. 
In this paper, we describe the design and development of this instrument, its validation, and measurements made in classes at CU Boulder and elsewhere.
\end{abstract}

\maketitle 

\section{Introduction}\label{sec:intro}

In recent years, the physics education research (PER) community has placed a strong emphasis on improving student learning in upper-division courses for physics majors.\cite{Manogue:2001cd,Singh:2001el,Chasteen:2012ku,Chasteen:2012cz} Many research studies have shown the wide variety in students' understanding of particular physics concepts and practices during and after instruction. \cite{Pepper:2012ka,Wallace:2010bb,2007AIPC..883..185S,Pollock:2007wv,Smith:2011tr,Wilcox:2015gb,Wilcox:2013ea,Turnbull:2015jf,Manogue:2006hy} Systematic efforts to assess student understanding on a broader scale have been useful in facilitating this effort.\cite{PhysRevSTPER.11.020115} These systematic assessments of student understanding at the upper-division highlight common and persistent student difficulties that can both inform curricular and pedagogical innovations and help form the basis for research efforts. Furthermore, these measures of student performance provide an indicator of the effectiveness of different pedagogies and curricula and can be used by instructors and departments to improve course offerings over time.

{In fact, over the last 40 years, the awareness created by assessments of student learning using conceptual inventories has helped to drive widespread transformation of introductory lecture courses.\cite{Hestenes:1992ws,Thornton:1998tk,Ding:2006kq}} The use of these conceptual inventories has also helped the physics community identify persistent difficulties and provided the means to compare learning outcomes between different pedagogical and curricular reforms as well as across many institutions and implementations.\cite{McDermott:1999tz,Hsu:2004kh,Meltzer:2012eg,Hake:1998cq,Kohlmyer:2009ib,Caballero:2012ee,ding2014uncovering,von2016secondary} 

Over the last decade, the Department of Physics at the University of Colorado Boulder (CU) has worked to transform their upper-division lecture courses to more student-centric instruction.\cite{Chasteen:2009wn,PhysRevSTPER.11.020110} This transformation process has involved the development of faculty-consensus learning goals,\cite{2012AIPC.1413..291P} the development of instructional materials,\cite{Pollock:2012uy} and research to identify student difficulties,\cite{Caballero:2012wr,Wilcox:2013ea,Turnbull:2015jf,Pepper:2012ka} which has informed refinements to both the aforementioned learning goals and instructional materials. In recent years, upper-level assessments in the areas of quantum mechanics \cite{sadaghiani2015quantum} and E\&M \cite{Chasteen:2012fl,wilcox2014coupled} have been developed to, in part, understand the impact {of these transformations on student understanding.}

In this paper, we present the Colorado Classical Mechanics/Math Methods Instrument (CCMI) that is both grounded in the history of this work and opens a new space for upper-level physics assessments -- middle-division Classical Mechanics and Mathematical Methods (CM 1).
The CCMI (Sec.~\ref{sec:ccmi}) consists {mostly of open-ended questions} that probe students' use of the sophisticated skills and practices outlined in faculty-consensus learning goals. 
In Sec.~\ref{sec:dev}, we present the development of the CCMI including the design of the questions and the measures that provide evidence of validity. We discuss the design and structure of the grading rubric as well as measures of reliability in Sec.~\ref{sec:grading}. In Sec.~\ref{sec:stats}, we present statistical results from its implementation at CU and other institutions through the lens of classical test theory. Finally in Sec.~\ref{sec:closing}, we discuss implementation, measurement, {and possible uses}.

\section{The Colorado Classical Mechanics/Math Methods Instrument}\label{sec:ccmi}

\begin{figure*}

\begin{minipage}{\linewidth}
\begin{mdframed}
\vspace*{5pt}
\flushleft {\bf Learning goals evaluated: {\it Students should be able to}}:\\
$\cdot$ choose appropriate area and volume elements to integrate over a given shape.\\
$\cdot$ translate the physical situation into an appropriate integral to calculate the gravitational force at a particular point away from some simple mass distribution.\\
\vspace*{-4pt}
\noindent\rule{\linewidth}{0.4pt}\\
\vspace*{5pt}
\begin{minipage}{0.72\linewidth}
\flushleft {\bf Q9} Consider an infinitely thin cylindrical shell with non-uniform mass per unit area of $\sigma (\phi, z)$. The shell has height $h$ and radius $a$, and is not enclosed at the top or bottom.\\
\flushleft (a) What is the area, $dA$, of the small dark gray patch of the shell which has height $dz$ and subtends an angle $d\phi$ as shown to the right?\\
(b) Write down (BUT DO NOT EVALUATE) an integral that would give you the MASS of the entire shell. Include the limits of integration.
\end{minipage}
\begin{minipage}{0.25\linewidth}
\flushright
\includegraphics[clip,trim=70mm 40mm 70mm 30mm,width=0.75\linewidth]{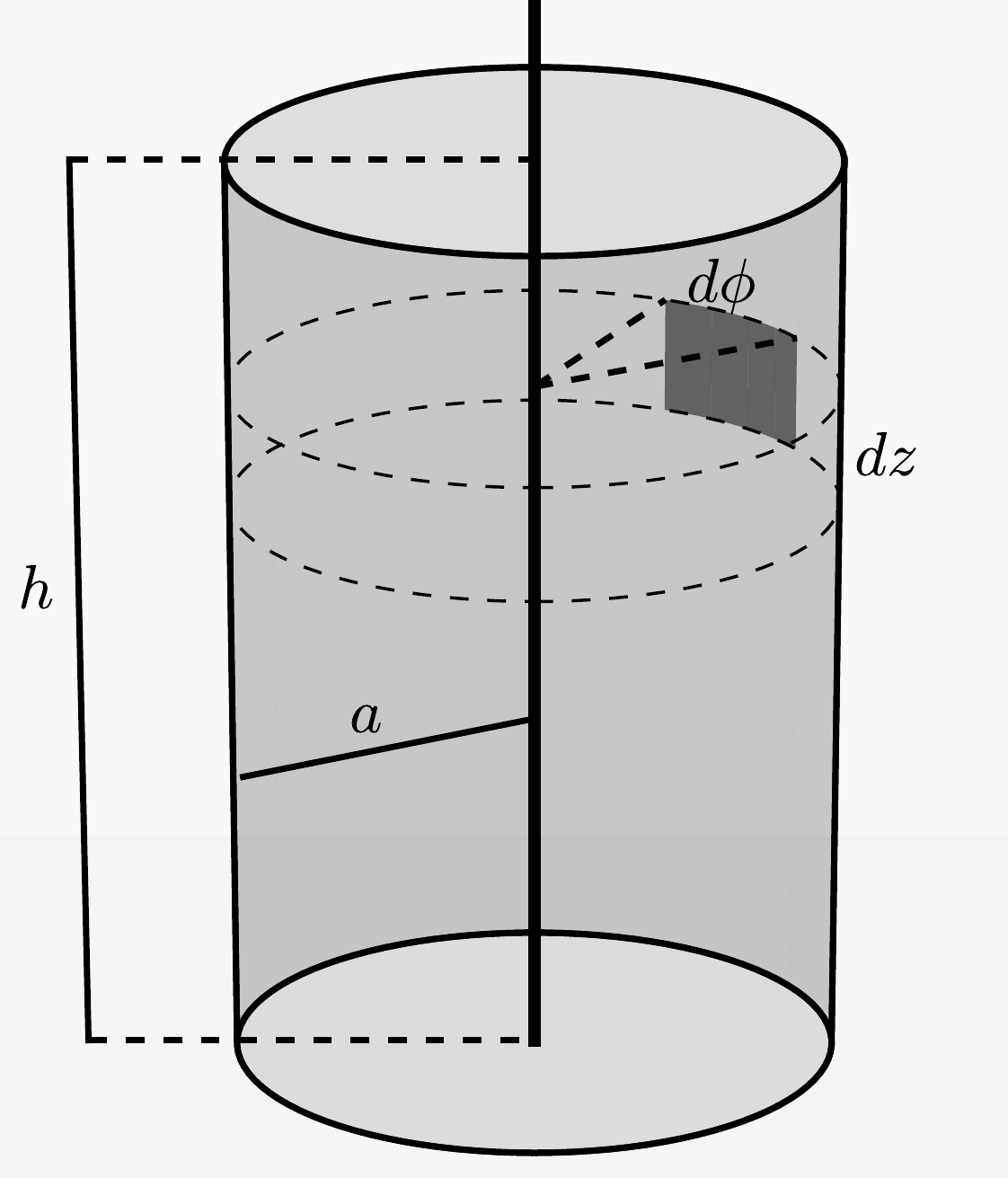}
\end{minipage}
\vspace*{5pt}
\end{mdframed}
\end{minipage}

\caption{Certain topic-scale learning goals are evaluated by the CCMI questions. The sample question appears on the CCMI pre- and post-tests; vector calculus is a prerequisite for CM 1. This question constitutes 9\% of the total post-test score.}\label{fig:mass}

\end{figure*}

The Colorado Classical Mechanics/Math Methods Instrument (CCMI) is a 9-question {(with a total of 22 sub-parts)} open-ended test that focuses on topics taught in the first half of a two-semester classical mechanics sequence. This first course concludes before a discussion of the calculus of variations; hence, the Lagrangian and Hamiltonian formulations of mechanics are absent from the test. The CCMI focuses on core skills and commonly encountered problems. Students solve a variety of problems such as:~determining the general solution to common differential equations (e.g., $\ddot{x}=-A^2x$); finding equilibria and sketching net forces on a potential energy contour map; and decomposing vectors in Cartesian and plane-polar coordinates. We have designed the CCMI to be given in a standard 50-minute lecture period. To accompany the longer post-test, we have developed a short (15-20 minute) pre-test that contains a subset of three problems {taken from the post-test.} Figure \ref{fig:mass} shows a sample CCMI question that appears on both the pre- and the post-test. Table \ref{tab:QuestionTable} contains {the full listing of questions on the CCMI.} The complete CCMI and the accompanying support documents are available online.\cite{CMweb}

\begin{table*}[h]

\begin{tabular}{p{1cm}p{0.7cm}p{3cm}p{7.2cm}p{1.5cm}p{1.2cm}}
\toprule

{\bf{Q no.}} & {\bf{Pts}} & {\bf{Short name}} & {\bf{Description}} & {\bf{Cohen's kappa}} & {\bf{Pearson coeff.}}\\

\hline 

Q1 & 3 & Common differential equations & Write the general solution to the differential equations $\ddot{x} = -A^2 x$ and $\frac{dy}{dt} = By$. & 0.42 & 0.54 \\

\hline

Q2 & 2 & Taylor approximation & Given $\Delta g = \frac{G\;M_E}{(R-d)^2} - \frac{G\;M_E}{R^2}$, explain how you would determine an approximate formula for $\Delta g$ if d is small. & 0.63 & 0.25 \\

\hline

Q3 & 5 & Potential energy map & Potential energy plot of a particle free to move on a 2-d plane. Where is the particle in stable equilibrium? Rank the magnitude of the gradient at points on the plot. Draw vectors that represent the force at those points & 0.75 & 0.50 \\

\hline

Q4 & 5 & Damped harmonic oscillator & Expression,$a_1\ddot{x}+a_2\dot{x}+a_3 x = 0$, and a corresponding graph for the motion of mass on a spring. Identify the units of $a_1$, $a_3$, and resketch the solution if $a_3$ is smaller. & 0.57 & 0.47 \\

\hline

Q5 & 3 & Simple harmonic oscillator & In simple harmonic motion, what is restoring force proportional to? Write an expression for position as a function of time. Draw potential energy as a function of position. & 0.88 & 0.59 \\

\hline

Q6 & 6 & Vector decomposition & Ball sliding in the bottom of a sawed off sphere. Draw the vectors $\hat{r}$ and $\hat{\theta}$. Express the velocity vector in the $x$--$y$ and $r$--$\theta$ coordinate systems.  & 0.45 & 0.56 \\

\hline

Q7 & 2 & Resonance & Mass on a frictionless spring attached to a driving force with a small amount of friction in the system. Sketch the amplitude of the oscillation of the mass as a function of the driving frequency. & 0.90 &  0.46 \\

\hline

Q8 & 4 & Writing a differential equation & Given description of the position and forces between a particle confined to move between two objects that attract it. Write down a differential equation that describes the position of the particle as a function of time. & 0.67 & 0.54   \\

\hline

Q9 & 3 & Writing an integral & Infinitely thin cylindrical shell with non-uniform mass per unit area. Write down an integral that would give you the mass of the entire shell. & 0.78 & 0.50 \\

\hline

Q10* & 2 & Fourier series & Which Fourier series could be the correct expansion for the given function? (MCQ) & - & - \\

\hline

Q11* & 1 & Laplace's equation & How would you separate $U$ to solve Laplace's equation in Cartesian coordinates, $\frac{\partial^2 U}{\partial x^2} + \frac{\partial^2 U}{\partial y^2}$ = 0, using separation of variables? (MCQ) & - & - \\

\hline
\hline

\end{tabular}\caption{Questions appearing on the CCMI. The full instrument is available online.\cite{CMweb} {Questions 10 and 11 are both optional (*) and multiple-choice questions (MCQ)}}

\label{tab:QuestionTable}
\end{table*}

\section{Development and Content Validation}\label{sec:dev}

{The development of {the CCMI} followed the process established by Chasteen et al., \cite{Chasteen:2012fl} which was recently reviewed by Wilcox et al. in their paper describing the uses and development of upper-level physics assessments.\cite{PhysRevSTPER.11.020115} Broadly speaking, the process involves establishing and prioritizing assessable learning goals, crafting questions that are tested with students using think-aloud interviews,\cite{Bowen:1994vh} and validating questions based on student and faculty input.}

\subsection{Development History}

At CU, CM 1 is a blended classical mechanics and mathematical methods course that forms the first half of a two-semester sequence in classical mechanics. In recent years, this course was transformed from lecture-based instruction to {more active and student-centric} instruction.\cite{PhysRevSTPER.11.020110} The early part of this transformation involved the development of consensus learning goals by a group of faculty. A series of faculty meetings were held to develop consensus course-scale learning goals and to articulate the topical content coverage of the course.\cite{2012AIPC.1413..291P} {Overall 19 faculty (4 PER, 15 non-PER) participated in at least one of the 7 meetings, with an average of 9 faculty at each meeting.\cite{2012AIPC.1413..291P}} {Course-scale learning goals focus on how the student develops over the whole semester. For example, students in CM 1 are consistently exposed to the connection between math and physics. Relevant course-scale learning goals for this area include: ``Students should be able to translate a physical description of a sophomore-level classical mechanics problem to a mathematical equation necessary to solve it. Students should be able to explain the physical meaning of the formal and/or mathematical formulation of and/or solution to a sophomore-level physics problem. Students should be able to achieve physical insight through the mathematics of a problem.''} 

After the development of course-scale learning goals, a set of specific, topic-scale learning goals were drafted. To develop these topic-scale learning goals, we utilized field notes collected during lectures, weekly homework help sessions, and faculty meetings. A further set of faculty meetings were held in which the topic-scale learning goals were agreed upon. In these meetings, several topic-scale learning goals were selected to be assessed on the CCMI as articulated in the learning goals.\cite{2012AIPC.1413..291P} 

These topic-scale learning goals combined content coverage that faculty had defined and the mathematical and problem-solving skills characteristic of upper-division coursework. For example, ``Students should be able to use Newton's laws to translate a given physical situation into a differential equation'' and ``Students should be able to project a given vector into components in multiple coordinate systems, and determine which coordinate system is most appropriate for a given problem.'' These course-scale and topical-scale learning goals are available online.\cite{CMweb}

These topic-scale (measurable) learning goals formed the basis for the development of the CCMI. While these learning goals were developed by CU faculty, and are specific to CM 1, we believe that many are applicable to the mathematical methods and classical mechanics courses offered at other universities because (1) the goals were developed by traditional{, not PER,} physics faculty, and (2) the topical coverage was drawn from {the first five chapters of} a standard classical mechanics textbook.\cite{taylor2005classical} Moreover, faculty from {five} other institutions have given the CCMI in their courses and were interviewed to obtain feedback on the learning goals assessed by the CCMI as well as the CCMI itself. These interviews led to changes in coverage and scoring of the CCMI.

As the topic-scale learning goals were developed, CU faculty discussed which ones were most fundamental to student learning, that is, which goals (when met by students) would be taken as evidence of learning in CM 1, which goals formed the basis for future learning (e.g., in future physics coursework), and, thus, which goals should be assessed on a standardized instrument. When a topic-scale goal was deemed by faculty to be assessment-worthy, a draft assessment item was written by the postdoctoral researcher facilitating these conversations with input from faculty. Sixteen open-ended questions were initially written.  Some of these questions were adapted from exam or clicker questions written by CU faculty in previous semesters. All questions were informed by observed student difficulties.\cite{Pollock:2012uy} The early versions of these questions were entirely open-ended and were developed to draw out student ideas about the particular concepts and skills that would be assessed on the {final instrument.} 

The earliest version of the CCMI contained 16 questions -- more than could be answered in a single 50-minute class period. Thus, the CCMI was split into two 11-question versions with some number of overlapping questions; each version was given to half the students in the class. One benefit of developing this instrument at a large, research-intensive university is a large population of students taking CM 1 -- in some semesters, more than 100 students have been enrolled in CM 1 at CU. Through a number of administrations of early versions of the questions, feedback from faculty and students, as well as timed testing, the CCMI was trimmed to an 11-question, open-ended assessment that could be administered in a 50-minute period. Following this internal development period, the CCMI was offered in a ``beta'' version to faculty teaching courses like CM 1 at other institutions. Administration of the CCMI at these other institutions provided additional feedback on the content coverage and scoring of the CCMI. 

Interviews were conducted with these ``beta'' testers to learn more about their courses, their needs, and their view of the CCMI. These interviews were prompted by concerns about certain questions from the Colorado Upper-division Electrostatics Assessment from colleagues using it at other institutions.\cite{zwolak2013re} Prior to these interviews, faculty were given a copy of the CCMI and the accompanying learning goals (e.g., Fig.~\ref{fig:mass}) to review. The ``CM 1'' courses that our interviewees taught ranged from quite similar to CM 1 (e.g., a 2 semester sequence classical mechanics) to quite compressed compared to CM 1 (e.g., a 1 semester course on classical mechanics that surveys all common topics including Lagrangian Hamiltonian dynamics and the orbit equation). While there was a substantial diversity among the topical coverage among the courses taught by these faculty, most agreed that 9 of the 11 questions were covered well enough in their courses to be included as part of the assessment. However, for most faculty, 2 questions, which deal with Fourier Series and Laplace's equation, were covered superficially or not at all in their courses. As a result, the CCMI consists of 11 questions -- 9 core questions that count towards the overall score that can be compared across institutions, and 2 optional questions that may be used at institutions where such topics are taught.

\subsection{{Content} Validation of the CCMI}

{In designing the CCMI, we took the approach that an assessment of student learning should address the topics that traditional physics faculty value. This serves to validate the instrument in the sense that the questions being asked of students cover the topics in the way that faculty desire. {Further, this process serves to generate buy-in to use the instrument.}} Secondly, the instrument needs to be interpretable by students, that is, students need to be able to interpret each question consistently and in the ways that instructors expect. Below, we detail how we established the validity of the CCMI through discussions with faculty and think-aloud interviews with students.

\paragraph{Expert Validation}

As the basis for the questions were the expert-developed learning goals,\cite{2012AIPC.1413..291P} the instrument was grounded in the topics deemed essential by faculty. Draft questions were developed from these learning goals; some were inspired by existing course materials (clicker questions, exam questions, etc.) and others were crafted from scratch. Once a complete set of questions was drafted, faculty at CU and elsewhere were consulted individually to obtain their feedback on the instrument. The CCMI was sent to faculty before meeting with the postdoctoral researcher for a semi-structured interview. The instrument and subsequent questions were positioned to the interviewed faculty in the following way: 
\begin{quote}
 ``Does this question ask about the kinds of things you want students in your CM 1 class to learn?''  

 ``If a student in your CM 1 class correctly solved this question, would you say that student demonstrated an understanding of this topic? Why?''  

 ``If a student performed well on this instrument, would you expect them to have performed well in your CM 1 class? Why?'' 
\end{quote}
As faculty spoke on these different topics, follow-up questions were asked to elucidate the meaning behind faculty's answers. In all, nine faculty {(4 at CU and 5 elsewhere; all non-PER faculty)} were interviewed for between 50-90 minutes. Individual faculty input was often aligned with each other, likely because these interviews took place following the discussion of learning goals. But, there were conflicting comments at times. For example, most faculty interviewed agreed that the instrument should focus on conceptual aspects of CM 1 while one or two faculty desired students to perform calculations on certain questions (e.g., Taylor series) because they believed that to be the only way to judge student learning on those particular topics. Where there was disagreement between interviewed faculty, we sided with the majority. Hence, the CCMI focuses on more conceptual aspects of CM 1. Faculty input was critical to deciding which questions to prune from the 16-question version of the CCMI. Discussion with faculty lead to ranking questions by ``most important for students to understand after completing CM 1.''

\paragraph{Student Validation}

Questions on the CCMI were further shaped by conducting think-aloud interviews with students while they solved the CCMI. The interviews served two purposes: (1) to ensure that the wording of the questions was clear for students (i.e., that students would {interpret questions as asked}), and (2) to collect student reasoning for correct and incorrect answers in order to help shape the grading rubric, which had not yet been fully designed. {Eight CU students who had recently completed CM 1 earning grades ranging from A to C} were interviewed {(in two cohorts)} for 60-90 minutes as they solved the CCMI. {Following a think-aloud protocol,\cite{fonteyn1993description} students narrated their thoughts while solving each question.} {The interviewer took notes identifying how each student read each question, what reasoning was brought to bear on each question, and where there were points of confusion or issues of clarity.} If at any time, the student struggled to answer a question, the interviewer suggested they make their best attempt given what they understand. Following a student's completion of the CCMI, the interviewer followed-up {question by question} with the student about their reading of the questions and their reasoning through their answer. The interviewer also discussed the correct solution to each question with most students as they were often interested in how well they performed. These interviews and notes were analyzed for salient themes that addressed issues of clarity and student reasoning after the first cohort of students completed the interviews. 

{The most prevalent issues were addressed by the first round of editing by the development team. For example, no interviewed student in the first cohort knew how to answer the Taylor series question. Discussion with the interviewees indicated a mismatch between our intent (i.e., explaining the importance of the small parameter in the expansion) and their experience (i.e., not ever being asked to think explicitly about the small parameter).} Questions were redrafted before conducting interviews with the next cohort of students. In this second set of interviews, the majority of questions elicited the expected responses and underlying reasoning. Those questions that still had some issues were positioned to the students as,

\begin{quote}
``In this question, we are trying to get you to work with this idea (e.g., Taylor series) in this way (e.g., identifying the small parameter in the expansion), how would you know to do that?''
\end{quote}

Students' responses to questions of this kind provided the final edits to the previously-problematic questions.

\begin{figure*}[t]

\begin{minipage}{\linewidth}
\begin{mdframed}
\vspace*{5pt}
\flushleft {\bf Question 9 (Writing an Integral) - Total points: 3}:\\
\vspace*{8pt}
Part A: 1 point\\
\begin{tabular}{p{0.2\linewidth}|p{0.2\linewidth}|p{0.5\linewidth}}\hline
Full credit (1) & Correct & $a d\phi dz$\\
Partial credit (0.75) & Wrong length scale & $r d\phi dz$\\
No credit (0) & Incorrect & No credit for any other responses\\\hline
\end{tabular}\\

\vspace*{8pt}
Part B: 2 points\\

\begin{tabular}{p{0.2\linewidth}|p{0.2\linewidth}|p{0.5\linewidth}}\hline

Full credit (2) & Correct & Correct integral form: 
\begin{itemize}[noitemsep]
\item Integrals must be over $d\phi$ and $dz$; $dA$ must agree with part A
\item Limits on the $d\phi$ integral must be $[0,2\pi]$
\item Limits on the $dz$ integral must be either $[0,h]$ or $[-h/2,h/2]$
\item Mass density ($\sigma(\phi,z)$ or $\sigma$) must appear in the integral; Substituting $\rho$ is OK
\end{itemize}\\

Partial credit (1.75) & ``Symmetric'' integral & Integral over $d\phi$ is from $[0,\pi]$ but is multiplied by 2\\

Minus 0.5 points (1 max) & Incorrect 2D integral & Problematic limits or kernel:\begin{itemize}[noitemsep]
\item Limits on the integral are incorrect (e.g., $[-h,h]$ or $[0,\pi]$)
\item Kernel of the integral is incorrect (e.g., missing $\sigma(\phi,z)$ or $a$)
\end{itemize}\\

No credit (0) & Incorrect & No credit for any other responses (e.g., 1D or 3D integral)\\\hline

\end{tabular}
\vspace*{8pt}
\end{mdframed}
\end{minipage}

\caption{Grading rubric for the question appearing in Fig. \ref{fig:mass}. The format focuses the grader's attention on the final response provided by the student. The grading rubric was not designed to elucidate details of student difficulties, but rather to capture the common final responses provided by students and score them accordingly.}\label{fig:massgrade}

\end{figure*}

\section{Grading the CCMI}\label{sec:grading}

{Scoring student responses to an assessment reliably undergird the value of the assessment to students and faculty. The rubric for the CCMI was informed by the lessons learned from our development of other upper-level assessments\cite{PhysRevSTPER.11.020115} as well as experienced and anticipated challenges for faculty users.}

\subsection{Rubric Rationale}

With the validity of the CCMI established, we turned to scoring student responses to provide an indication of student performance. It is important for independent assessments of student learning, such as the CCMI, that independent graders achieve consistent results. Therefore, the scoring rubric needs to capture the variety of student responses and indicate how each response is scored. There are a number of possible approaches to supporting graders in this work. {For example, in the electrostatics context, the Colorado Upper-Division Electrostatics Diagnostic (CUE)} took the approach of training graders to attend to both students' final answers and the nuances of student responses.\cite{Chasteen:2012fl} As such, graders were not only providing a consistent score for student work, but also attending to the details of student difficulties. The training was not intended to be overly prohibitive ($\sim$8 hours), but there was not much interest outside the PER community to learn to grade the CUE. Thus, researchers at CU have continued to provide a grading service to the physics community. In order to facilitate grading and promote wider use of the CUE, Wilcox, et al. developed a multiple-choice version of the CUE that can be delivered online.\cite{wilcox2014coupled} This work leveraged the large body of CUE responses collected over the years to develop an updated set of questions and a logical grading model that has proven quite successful - reproducing similar results {to the original CUE.}

While the CCMI has recently had significant interest from faculty at a number of institutions, the initial work to develop the rubric could not leverage a large body of responses. Thus, we decided to separate the two roles of the assessment into two rubrics: (1) a grading rubric that allows for scoring student work from a ``mastery'' perspective \cite{Caballero:2014dd,2012arXiv1207.6040B} and (2) a difficulties rubric that helps to uncover the prevalence of student difficulties in CM 1.\cite{Doughty:2015gs} The grading rubric is intended for faculty with no training to grade their students' responses consistently and have confidence that their scoring is meaningful. The difficulties rubric {that we are developing} is intended for researchers (or faculty) who intend to dig deeper into student reasoning and requires some amount of training. In this paper, we discuss only the grading rubric. 

The approach to grading the CCMI that we have used focuses on the students' final answer and points are taken away for errors in that answer. Graders need only to attend to one part of a student's answer and can score based on more salient features of the student's final response. This grading approach is taken by both the CCMI and {the Colorado UppeR-division ElectrodyNamics Test (CURrENT).} \cite{2012arXiv1207.6040B} The development of the grading rubric was grounded in patterns in students' responses to CCMI questions, which formed the basis for categories in the grading rubric.\cite{Doughty:2015gs,Turnbull:2015jf} The grading rubric describes how points are deducted for different errors, providing examples where necessary (it does not list all the possibilities). The illustrative errors are those commonly seen in students’ answers. The allocation of points on each question and the partial credit awarded for some responses are based on faculty ``rankings'' of the relative importance of the learning goal each question assesses and the relative importance of the errors. The rubric used to grade the question shown in Fig. \ref{fig:mass} appears as Fig. \ref{fig:massgrade}. Large-scale (N$>$500) use of the rubric on students' responses at CU and other institutions resulted in changes to the wording of the rubric and the addition of new examples. While a different design for scoring student work might be used, in our design, we considered asking traditional faculty to grade the assessment and how we might achieve consistent results across untrained graders. Our grading procedure does produce consistent results. 

\subsubsection{Inter-grader Reliability}

Through a series of analyses, we established the reliability of our grading rubric. Our work follows the analysis conducted by Chasteen, et al. to establish a reliable grading rubric for the CUE, \cite{Chasteen:2012fl} but also makes use of an untrained grader who was asked to use the completed rubric to score student responses. The two graders (one untrained) scored responses from 100 students to all 11 questions on the CCMI. The resulting scores assigned to individual responses were compared as well as the overall score for a given students' CCMI. The resulting analysis demonstrated that an untrained grader can score students' responses to the CCMI reliably using the grading rubric.

\begin{figure}
\includegraphics[width=\linewidth]{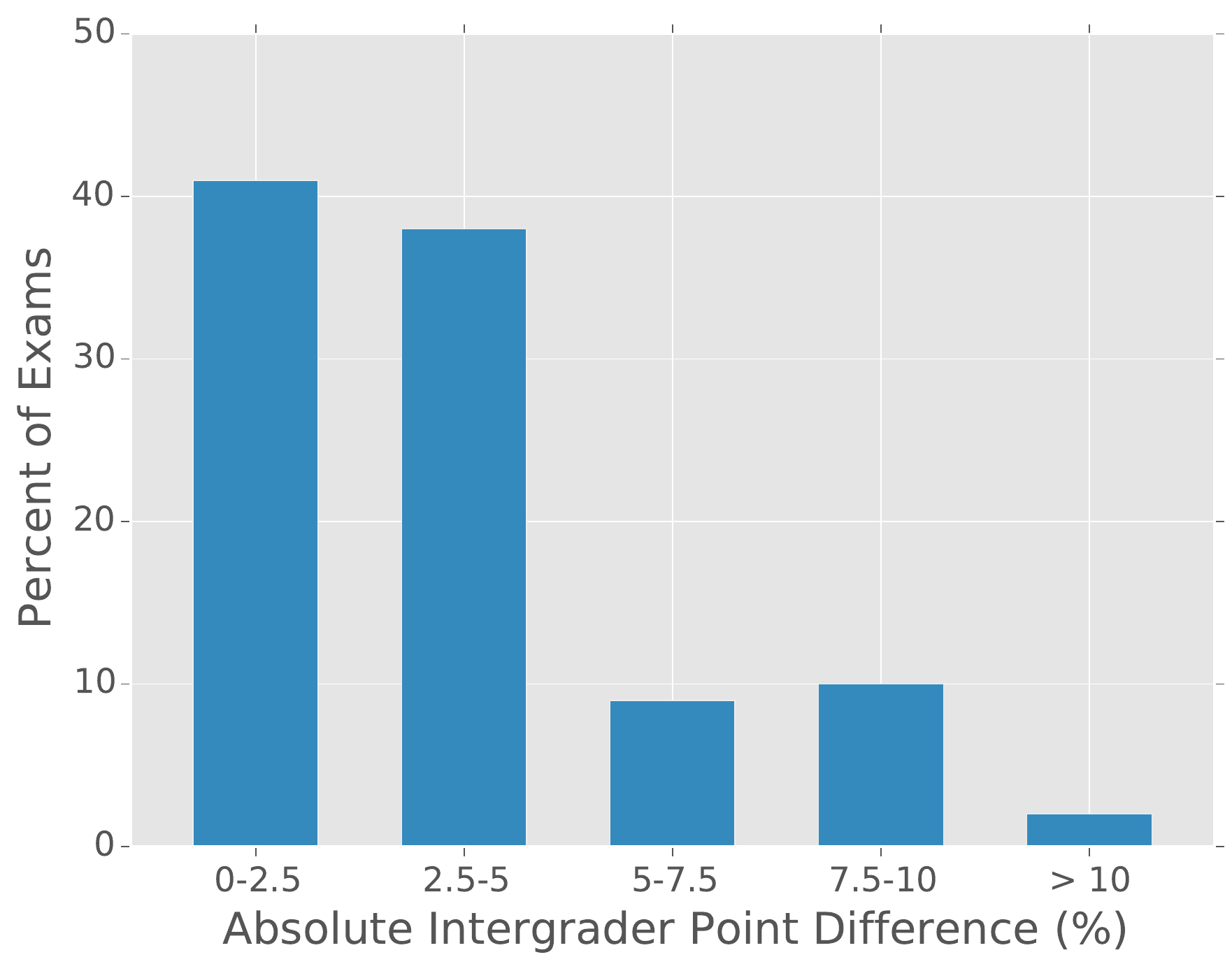}
\caption{[Color online] The absolute difference in CCMI scores assigned by a trained and untrained grader is presented. The average difference on total CCMI score between the trained and untrained grader is 3.5\% $\pm$ 2.7\%. The graders agreed to 5\% on overall score for 79\% of the exams.}\label{fig:absdiff}
\end{figure}

First, the average overall difference in CCMI scores assigned to students between a trained and untrained grader is less than 5\% (3.5\% $\pm$ 2.7\%) of the total points. Fig. \ref{fig:absdiff} demonstrates that the graders agreed on a total score within 10\% for all but two students, and for the majority of students (79\%) the graders were below 5\% disagreement.

\begin{figure*}
\includegraphics[width=\linewidth]{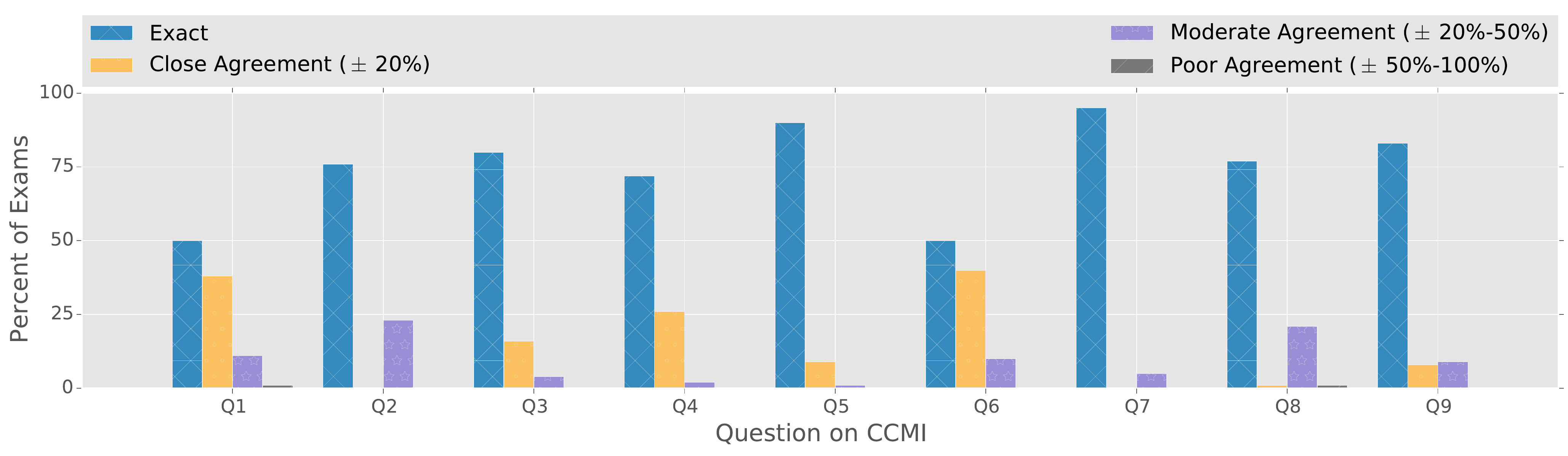}
\caption{[Color online] For each question, the score differences between a trained and untrained grader are shown. These percentage agreement between graders was binned as exact (same score, blue), close (within $\pm$ 20\%, orange), moderate (within $\pm$ 20\%-50\%, purple), and poor (more than $\pm$ 50\%, gray) agreement. For all questions, exact agreement was the most prevalent form of agreement ($\geq$50\% of exams for all questions).}\label{fig:reliability}
\end{figure*}

While this difference on total score is an intuitive measure of agreement, a more rigorous test of agreement is one that includes where graders agree by chance. 
It is typical in assessment work to use Cohen's Kappa to characterize the agreement between two (or more) graders.\cite{Cohen:1960tf,cohen1968weighted} 
However, there are concerns with using Cohen's Kappa were partial credit is awarded, where the scales between items differ, and where the total number of possible scores is high. 
Moreover, Cohen's Kappa is a relatively conservative measure of agreement. \cite{brennan1981coefficient} 
These issues associated with Cohen's Kappa for determining reliability on an assessment like the CCMI appears in Chasteen, et al.\cite{Chasteen:2012fl} 
As expected (and previously observed), agreement across all possible point distributions is low ($\kappa$ = 0.23). 
It is unlikely for each grader to agree on the overall points awarded to each student (Fig. \ref{fig:absdiff}), but it is fairly likely for graders to agree within a few points. 
Like the CUE, Cohen's Kappa calculated for scored binned into two-point intervals ($\sim$5\%) provides evidence of moderate agreement ($\kappa$ = 0.47). 
When binned into four-point intervals ($\sim$10\%), we obtain evidence of substantial agreement ($\kappa$ = 0.64). 
Hence, within differences of 5\%, we find that our trained and untrained graders agree well.

While this overall agreement is reasonable, it may be that specific questions may contribute to these differences more than others. That is, it might be that some combination of a specific question and the rubric describing how to score that question is unreliable. By determining Cohen's Kappa for each question on the CCMI (see Table \ref{tab:QuestionTable}), we find some evidence that questions 1 (Common differential questions) and 6 (Vector decomposition) might be contributing to these overall discrepancies. This message is further bolstered by the evidence provided in Fig. \ref{fig:reliability} where we have shown the percent agreement between a trained and untrained grader on each question. Here, we define ``exact'' to be the same score for the students' response while ``close'', ``moderate'', and ``poor'' represent agreement to $\pm$ 20\%, $\pm$ 20\%-50\%, and $\pm$ 50-100\% respectively.

These analyses provide evidence of a robust and reliable grading rubric, but we acknowledge that due to our design there is some information lost, particularly if the CCMI rubric is compared to the CUE rubric. Due to the focus on final answers, information about student difficulties that would be captured in a more detailed rubric is lost. We are developing a separate difficulties rubric to address this issue.\cite{Doughty:2015gs} However, what is gained (speed, accuracy, and adoption) by this approach to grading should not be understated.

\section{Statistical Validation of the CCMI}\label{sec:stats}

To establish an assessment as a valid and reliable instrument, further analysis into specific properties of the test must be conducted. Recently, this kind of work has shifted towards using response modeling techniques such as Item Response Theory (IRT).\cite{lord1980applications,embretson2013item} While IRT is quite robust and used widely, the body of data needed to use it reliably is more than we have been able to collect. Over the last several years, we have collected data from five CM 1 courses at CU (N = 244) and from eleven similar courses at nine other institutions (N = 218). There are simply not as many users or students taking upper-level assessments of this type. Hence, we make use of Classical Test Theory\cite{crocker1986introduction} -- following the analysis conducted by Chasteen, et al.\cite{Chasteen:2012fl} and Wilcox et al.\cite{wilcox2014coupled}

\paragraph{Internal Consistency} An assessment of student learning should be internally consistent. If the assessment aligns with the goals of instruction, students who perform well on a single question should perform well on other questions. Essentially, each question should provide consistent information about a student performance (on the average). It is typical to use Cronbach's alpha to investigate internal consistency {-- estimating the reliability of scores or the ``unidimensionality'' of the assessment.} We determined Cronbach's alpha treating each part of a question as an item because the total number of test items on CCMI is small. We find that the CCMI is a highly internally reliable assessment ($\alpha$ = 0.83). {The acceptable range for $\alpha$ is above 0.7, with greater than 0.8 being ``good.''\cite{cronbach1951coefficient}}

\paragraph{Criterion Validity} If aligned well with the learning goals for a course, we expect that an independent assessment of student learning (i.e., the CCMI) should correlate with other assessments of student learning (e.g., final exams).  Students' exams are the most similar measure to the CCMI. Like exams, the CCMI is completed individually in timed and controlled environments. But, unlike exams, it does not affect students' grades. Each class at CU took three exams: two regular hour exams and a final. The averages of those three exams were {normalized (using a z-score, $z=\frac{x-\bar{x}}{\sigma}$)} to allow
comparisons of different instructors. CCMI post-test scores were strongly correlated with these z-scored exam averages (r = 0.71); a linear model can thus account for 50\% of the variance in exam scores associated with CCMI scores. Similarly high correlations were observed on the CUE.\cite{Chasteen:2012fl}

\paragraph{Item-test correlation} We expect that the performance on individual items to correlate well with the overall score on the instrument (minus the item being tested). This correlation is expected from the premise that the whole assessment is a measure of a large construct -- knowledge of CM 1 concepts -- and that construct has underlying features -- e.g., Taylor series -- that will be more or less learned in similar amounts. We use Pearson's $r$ (linear correlation) to determine how well each item connects to the rest of the CCMI (see Table \ref{tab:QuestionTable} for values of $r$ for each item). We find that all items are above the established threshold ($r\sim0.2$) for item-test correlation. However, we note that Question 2 (Taylor series) correlates much less than the rest of the items do with the whole instrument. 

\begin{figure}
\includegraphics[width=\linewidth]{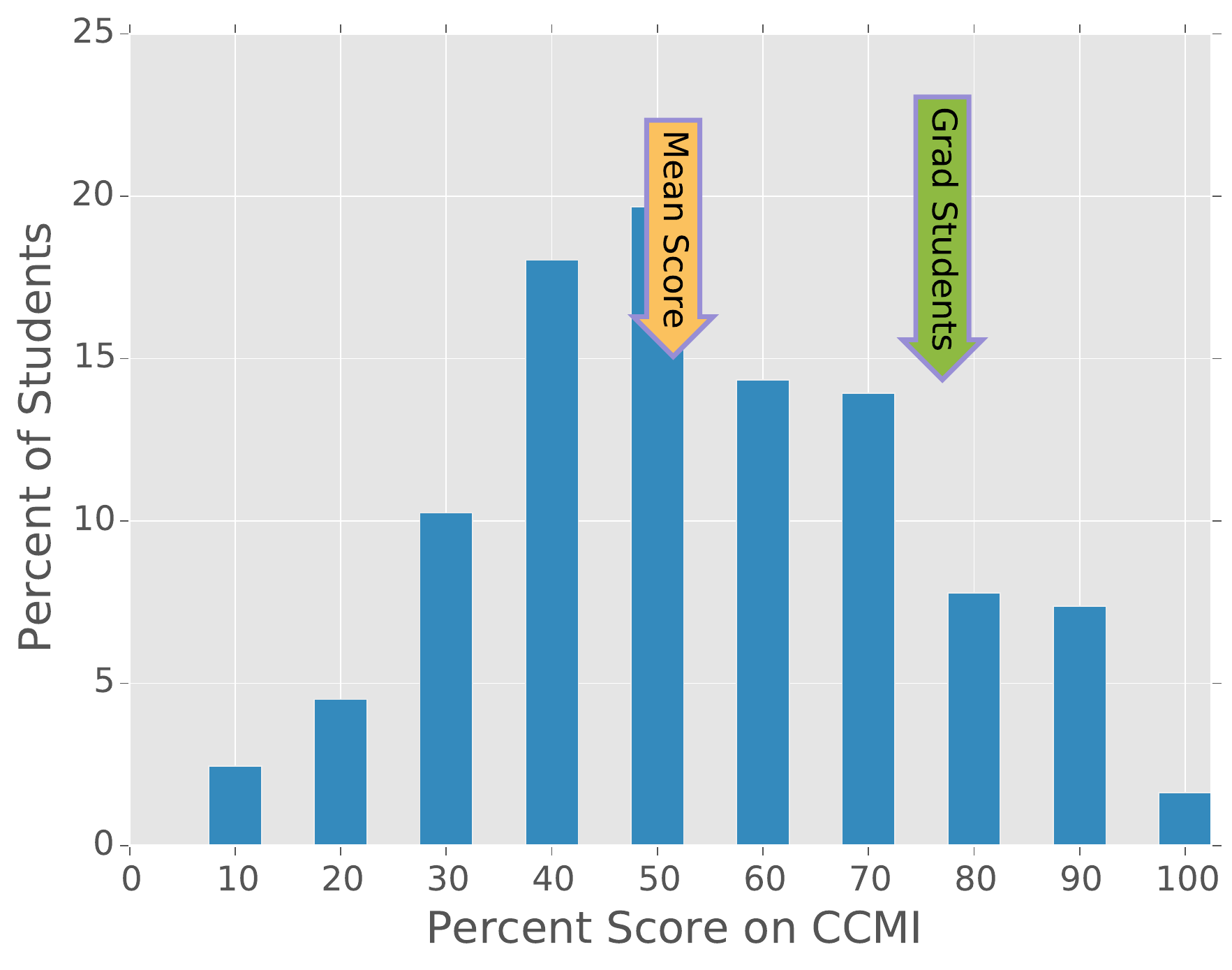}
\caption{[Color online] A histogram of student scores on the CCMI is presented (N = 462). The average score for students taking the CCMI is indicated (orange arrow): 49.0\% $\pm$ 1.0\% as well as the performance by first-year physics graduate students at CU Boulder (green arrow): 74.5\% $\pm$ 3.4\% (N = 5).}\label{fig:difficulty}
\end{figure}

\paragraph{Discrimination} An assessment of student learning should be able to separate students who demonstrated low understanding from those who demonstrated high understanding. Ferguson's delta is the typical measure of discrimination used for assessments of this type {-- it provides a measure of how broadly the scores are distributed across the possible scores.} In calculating Ferguson's delta, we used the total number of points on the assessment rather than the number of items as each question is worth a different number of points. We find that the CCMI has excellent discrimination on a per-point basis ($\delta = 0.99$). {A test with $\delta > 0.9$ is considered to have good discrimination.\cite{kline2013handbook}}

\begin{figure*}
\includegraphics[width=1\linewidth]{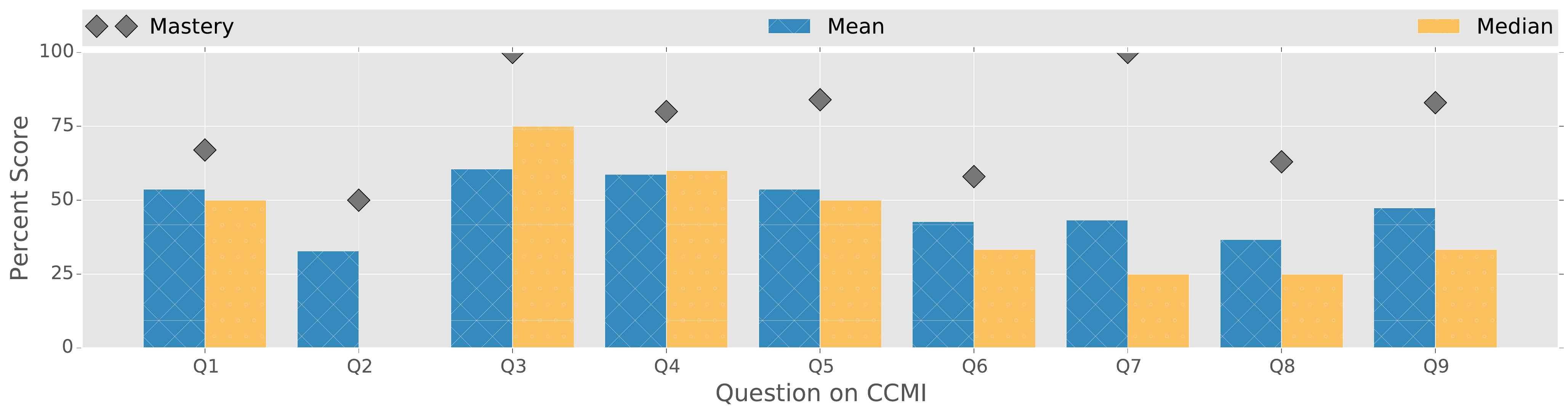}
\caption{[Color online] The mean (blue) and median (orange) of student performance on each question is provided. The median score for question 2 is zero. In addition, we show the mastery score (gray square): the score that separates the top three deciles from the bottom seven deciles, that is, the score which the top 30\% of scores lies above.}\label{fig:performance}
\end{figure*}

\paragraph{Additional analyses of discrimination} While Ferguson's delta is a typical measure, it might not be an intuitive measure of discrimination. In Fig. \ref{fig:difficulty}, we provide the histogram of student performance on the CCMI, which shows the mean score to be 49.0\% $\pm$ 1.0\% (N = 462). Indeed, the CCMI is a difficult assessment. First-year graduate students at CU earned an average score of 74.5\% $\pm$ 3.4\% (N = 5). In Fig. \ref{fig:performance}, we provide a visualization of the difficulty of each item. The mean and median score for each item are plotted along with the mastery score, {which is the score that the top 30\% of scores lies at or above.}

\section{Discussion and Conclusion}\label{sec:closing}

In summary, we have developed an assessment for Classical Mechanics/Mathematical Methods courses for which we have established validity and developed a reliable grading rubric. 
Scores on the CCMI correlate well with other measures of student understanding (i.e., in-class exams) and internal measures of validity, reliability, and discrimination are well within the acceptable scope for such an assessment.
While it may appear that the instrument is quite specialized, use by and feedback from faculty at other institutions have shaped the assessment to cover a broad range of courses, from those quite similar to CM 1 (e.g., a 2 semester sequence classical mechanics) to quite compressed compared to CM 1 (e.g., a 1 semester course on classical mechanics that surveys all common topics including Lagrangian Hamiltonian dynamics and the orbit equation).
That feedback from faculty informed both the design and use of rubric developed to analyze student work on the CCMI.
The design of the rubric for the CCMI separated the two traditional roles of assessment in physics education: (1) gaining a reliable understanding of student performance on specific topics, and (2) identifying persistent student difficulties.\cite{PhysRevSTPER.11.020115} 
The former role was presented in this paper as the grading rubric, which demonstrated reliability even when used by an untrained grader. 
A rubric to address the latter is in development and will be the subject of future work.

The CCMI was designed to serve a variety of purposes. 
Most simply, it is an independent measure of student understanding after instruction in a classical mechanics course.
Student performance on specific topics as well as performance across the instrument can serve {as a secondary and standardized measure of student understanding after a classical mechanics course.}
These measures can be used by faculty to improve different aspects of their instruction as they see fit.
Most faculty who have used the CCMI have used it for this purpose. 
Faculty have reviewed their score reports to identify strengths and weaknesses in their instruction based on their interpretation of students' scores {as well as to provide direct feedback to their students.}

At a slightly higher scale, the CCMI may serve as a tool for departments looking to assess their physics program.
It is becoming increasingly incumbent upon physics departments to demonstrate some form of independent assessment, and the CCMI (along with other standardized instruments) can help serve this purpose.
Unlike course final exams, the CCMI is a standardized instrument, which invites comparison over time, between curricula, and across institutions.
As such, student performance on the CCMI could be part of a more comprehensive departmental assessment.

From a cultural perspective, the CCMI offers opportunities for new (and seasoned) faculty to push on norms for teaching evaluation in their own tenure and promotion cases. 
Faculty teaching classical mechanics courses can demonstrate their commitment to quality instruction {by including student performance on CCMI in their teaching portfolios.}
These kinds of independent assessments are critical to understanding how student learning is being affected by instruction beyond the typical collection of course syllabi and student responses to end-of-course evaluations.

While we have developed a valid and reliable assessment for classical mechanics that can serve a number of purposes, we have accepted certain limiting factors in our design.
Given the constraints of administration (i.e., a 50 minute lecture period), the content coverage of the CCMI is limited (Table \ref{tab:QuestionTable}).
Not every instructor will agree on which topics should appear on an assessment for classical mechanics -- making it impossible to satisfy each instructor's needs. 
To address the issue of topical coverage, we drew from consensus learning goals\cite{2012AIPC.1413..291P} that were developed by traditional physics faculty.
In designing the questions for the CCMI, we worked with these faculty to prioritize the learning goals and, thus, the topics that were evaluated on the CCMI.
Furthermore, we collected feedback from instructors across the country to ensure that the CCMI meet most of their needs. 
It was in this work that two questions on the CCMI were designated optional as these topics were not covered to the degree they were at CU.
In a sense, we have developed an assessment {that serves as the ``common denominator''} for many implementations of classical mechanics.

A second limitation is our focus on students' final answers for the grading rubric, which under-emphasizes the process by which the student obtained the answer, and, moreover, can make it difficult to judge the prevalence of specific student difficulties.
The purpose of this answer-focused grading rubric was to streamline the process by which faculty can obtain information on student performance on the CCMI.
For example, a significant challenge for the CUE has been to train new graders to reliably score student responses to the CUE, which informed our decision to simplify the process so that an untrained grader using the rubric could score student responses reliably and have confidence that they had done so (Figs.~\ref{fig:absdiff} \& \ref{fig:reliability}).
Our current grading rubric has achieved this.

{To deal with this limitation, we are developing a rubric that helps categorize difficulties that manifest on the CCMI.\cite{Doughty:2015gs} This rubric is being informed by research into student understanding of classical mechanics.\cite{Wilcox:2013ea,Turnbull:2015jf}}
However, it is worth noting that there is still much that can be learned from scoring the CCMI as we have done: the most prevalent incorrect answers are represented in the grading rubric as partially correct answers (Fig.~\ref{fig:massgrade}).
In fact, our research into student's approaches to vector decomposition\cite{Turnbull:2015jf} was informed by results from grading the vector decomposition problem on the CCMI.
Hence, some information about the prevalence of certain kinds of student difficulties are captured by the grading rubric.
Wilcox et al. solved the problem of reliably scoring an independent assessment differently by adapting the CUE to a multiple-choice version with a logical scoring system {that could be offered online or with scantrons.}\cite{wilcox2014coupled}
This work benefited from the large body of student responses to the CUE collected over the years.
Now that we have completed the development of the CCMI and collected a similarly large body of student responses, we are exploring the possibility that the CCMI might be adapted into a {multiple-choice, machine-gradeable} format.

Our primary goal for developing the CCMI was to provide a tool for instructors to recognize {possible gaps between their instruction} and what students learn from that instruction. Our aim is to help instructors to adapt their teaching to align more with their own goals. As it exists presently, the CCMI can serve (and has already served) that purpose for a number of classical mechanics instructors.

\acknowledgments

We gratefully acknowledge the generous contributions of CU faculty, especially A.D. Marino, J.L. Bohn, K.P. McElroy, and collaborating faculty elsewhere. Particular thanks to the members of PER@C and PERL@MSU who have provided feedback and guidance on this work over the years. We also greatly appreciate the help of our student participants. This work was supported by University of Colorado’s Science Education Initiative and Michigan State University's College of Natural Science.

\bibliography{ccmi} 
\bibliographystyle{apsper}

\end{document}